\documentclass[aps,prl,twocolumn,preprintnumbers,showpacs,superscriptaddress,amsmath,amssymb]{revtex4}
\usepackage{graphicx} % Include figure files

\usepackage{epsfig}
\usepackage{color} % color
\begin{document}
%\doubles

\vspace*{-3\baselineskip}
\resizebox{!}{3cm}{\includegraphics{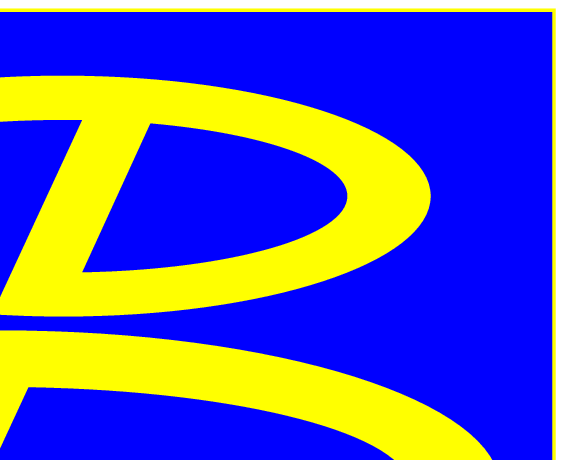}}

\preprint{\vbox{ \hbox{   }
%                 \hbox{   }
%                 \hbox{BELLE PAPER DRAFT}
%                 \hbox{Contact: S.~Uehara}
%                 \hbox{Internal Committee:}
%                 \hbox{\ \ R.~Mussa, S.~Olsen, B.~Yabsley}
%                 \hbox{ The first Version (V1.0) } April 8, 2009  }
%                 \hbox{\bf V6.3 Dec.18,~2009 }
                  \hbox{\large Belle Preprint 2009-29}
                 \hbox{\large KEK Preprint 2009-38 }
                 \hbox{December 2009}
                 \hbox{\ (February 2010, revised)}
}}

%\preprint{\vbox{ \hbox{ BELLE-CONF-00nn, Osaka nn \hfill}}}
%\twocolumn[\hsize\textwidth\columnwidth\hsize\csname] 
%@twocolumnfalse\endcsname 

\title{Observation of a charmonium-like enhancement in the
\boldmath{$\gamma \gamma \to \omega J/\psi$ } process}
\ \\

\normalsize
%%% Paper:    gamma gamma -> omega J/psi
%%% Journal:  Physical Review Letters
%%% Contacts: S. Uehara (uehara@post.kek.jp)
%%% Non-responding authors or those who said NO are commented out.
%%% ====================================================================
%%% Click the RELOAD button on your web browser to see the updated file.
%%% ====================================================================
%%% Use \input{author} to insert this material into your latex file.
%%%%% Force institutions to appear in alphabetical order when typeset.
\affiliation{Budker Institute of Nuclear Physics, Novosibirsk}
\affiliation{Faculty of Mathematics and Physics, Charles University, Prague}
%%%\affiliation{Chiba University, Chiba}
\affiliation{University of Cincinnati, Cincinnati, Ohio 45221}
%%%\affiliation{Department of Physics, Fu Jen Catholic University, Taipei}
\affiliation{Justus-Liebig-Universit\"at Gie\ss{}en, Gie\ss{}en}
\affiliation{The Graduate University for Advanced Studies, Hayama}
\affiliation{Gyeongsang National University, Chinju}
\affiliation{Hanyang University, Seoul}
\affiliation{University of Hawaii, Honolulu, Hawaii 96822}
\affiliation{High Energy Accelerator Research Organization (KEK), Tsukuba}
%%%\affiliation{Hiroshima Institute of Technology, Hiroshima}
%%%\affiliation{University of Illinois at Urbana-Champaign, Urbana, Illinois 61801}
\affiliation{Institute of High Energy Physics, Chinese Academy of Sciences, Beijing}
\affiliation{Institute of High Energy Physics, Vienna}
\affiliation{Institute of High Energy Physics, Protvino}
%%%\affiliation{Institute of Mathematical Sciences, Chennai}
\affiliation{INFN - Sezione di Torino, Torino}
\affiliation{Institute for Theoretical and Experimental Physics, Moscow}
\affiliation{J. Stefan Institute, Ljubljana}
\affiliation{Kanagawa University, Yokohama}
\affiliation{Institut f\"ur Experimentelle Kernphysik, Karlsruhe Institut f\"ur Technologie, Karlsruhe}
\affiliation{Korea University, Seoul}
%%%\affiliation{Kyoto University, Kyoto}
\affiliation{Kyungpook National University, Taegu}
\affiliation{\'Ecole Polytechnique F\'ed\'erale de Lausanne (EPFL), Lausanne}
\affiliation{Faculty of Mathematics and Physics, University of Ljubljana, Ljubljana}
\affiliation{University of Maribor, Maribor}
\affiliation{Max-Planck-Institut f\"ur Physik, M\"unchen}
\affiliation{University of Melbourne, School of Physics, Victoria 3010}
\affiliation{Nagoya University, Nagoya}
%%%\affiliation{Nara University of Education, Nara}
\affiliation{Nara Women's University, Nara}
\affiliation{National Central University, Chung-li}
\affiliation{National United University, Miao Li}
\affiliation{Department of Physics, National Taiwan University, Taipei}
\affiliation{H. Niewodniczanski Institute of Nuclear Physics, Krakow}
\affiliation{Nippon Dental University, Niigata}
\affiliation{Niigata University, Niigata}
%%%\affiliation{University of Nova Gorica, Nova Gorica}
\affiliation{Novosibirsk State University, Novosibirsk}
\affiliation{Osaka City University, Osaka}
%%%\affiliation{Osaka University, Osaka}
\affiliation{Panjab University, Chandigarh}
%%%\affiliation{Peking University, Beijing}
%%%\affiliation{Princeton University, Princeton, New Jersey 08544}
%%%\affiliation{RIKEN BNL Research Center, Upton, New York 11973}
%%%\affiliation{Saga University, Saga}
\affiliation{University of Science and Technology of China, Hefei}
\affiliation{Seoul National University, Seoul}
%%%\affiliation{Shinshu University, Nagano}
\affiliation{Sungkyunkwan University, Suwon}
\affiliation{School of Physics, University of Sydney, NSW 2006}
\affiliation{Tata Institute of Fundamental Research, Mumbai}
\affiliation{Excellence Cluster Universe, Technische Universit\"at M\"unchen, Garching}
\affiliation{Toho University, Funabashi}
\affiliation{Tohoku Gakuin University, Tagajo}
%%%\affiliation{Tohoku University, Sendai}
\affiliation{Department of Physics, University of Tokyo, Tokyo}
%%%\affiliation{Tokyo Institute of Technology, Tokyo}
\affiliation{Tokyo Metropolitan University, Tokyo}
\affiliation{Tokyo University of Agriculture and Technology, Tokyo}
%%%\affiliation{Toyama National College of Maritime Technology, Toyama}
\affiliation{IPNAS, Virginia Polytechnic Institute and State University, Blacksburg, Virginia 24061}
\affiliation{Yonsei University, Seoul}
  \author{S.~Uehara}\affiliation{High Energy Accelerator Research Organization (KEK), Tsukuba} % KEK
% \author{I.~Adachi}\affiliation{High Energy Accelerator Research Organization (KEK), Tsukuba} % KEK
% \author{H.~Aihara}\affiliation{Department of Physics, University of Tokyo, Tokyo} % Tokyo
% \author{K.~Arinstein}\affiliation{Budker Institute of Nuclear Physics, Novosibirsk}\affiliation{Novosibirsk State University, Novosibirsk} % BINP
% \author{T.~Aso}\affiliation{Toyama National College of Maritime Technology, Toyama} % Toyama
% \author{V.~Aulchenko}\affiliation{Budker Institute of Nuclear Physics, Novosibirsk}\affiliation{Novosibirsk State University, Novosibirsk} % BINP
  \author{T.~Aushev}\affiliation{\'Ecole Polytechnique F\'ed\'erale de Lausanne (EPFL), Lausanne}\affiliation{Institute for Theoretical and Experimental Physics, Moscow} % ITEP
% \author{T.~Aziz}\affiliation{Tata Institute of Fundamental Research, Mumbai} % Tata
% \author{S.~Bahinipati}\affiliation{University of Cincinnati, Cincinnati, Ohio 45221} % Cincinnati
  \author{A.~M.~Bakich}\affiliation{School of Physics, University of Sydney, NSW 2006} % Sydney
% \author{V.~Balagura}\affiliation{Institute for Theoretical and Experimental Physics, Moscow} % ITEP
% \author{Y.~Ban}\affiliation{Peking University, Beijing} % Peking
% \author{E.~Barberio}\affiliation{University of Melbourne, School of Physics, Victoria 3010} % Melbourne
% \author{A.~Bay}\affiliation{\'Ecole Polytechnique F\'ed\'erale de Lausanne (EPFL), Lausanne} % Lausanne
% \author{I.~Bedny}\affiliation{Budker Institute of Nuclear Physics, Novosibirsk}\affiliation{Novosibirsk State University, Novosibirsk} % BINP
  \author{K.~Belous}\affiliation{Institute of High Energy Physics, Protvino} % Protvino
  \author{V.~Bhardwaj}\affiliation{Panjab University, Chandigarh} % Panjab
  \author{M.~Bischofberger}\affiliation{Nara Women's University, Nara} % Nara
% \author{S.~Blyth}\affiliation{National United University, Miao Li} % NUU
% \author{A.~Bondar}\affiliation{Budker Institute of Nuclear Physics, Novosibirsk}\affiliation{Novosibirsk State University, Novosibirsk} % BINP
% \author{A.~Bozek}\affiliation{H. Niewodniczanski Institute of Nuclear Physics, Krakow} % Krakow
  \author{M.~Bra\v cko}\affiliation{University of Maribor, Maribor}\affiliation{J. Stefan Institute, Ljubljana} % Ljubljana
% \author{J.~Brodzicka}\affiliation{H. Niewodniczanski Institute of Nuclear Physics, Krakow} % Krakow
  \author{T.~E.~Browder}\affiliation{University of Hawaii, Honolulu, Hawaii 96822} % Hawaii
% \author{M.-C.~Chang}\affiliation{Department of Physics, Fu Jen Catholic University, Taipei} % FuJen
  \author{P.~Chang}\affiliation{Department of Physics, National Taiwan University, Taipei} % Taiwan
% \author{Y.-W.~Chang}\affiliation{Department of Physics, National Taiwan University, Taipei} % Taiwan
% \author{Y.~Chao}\affiliation{Department of Physics, National Taiwan University, Taipei} % Taiwan
  \author{A.~Chen}\affiliation{National Central University, Chung-li} % NCU
% \author{K.-F.~Chen}\affiliation{Department of Physics, National Taiwan University, Taipei} % Taiwan
% ...
% BINP
\collaboration{The Belle Collaboration}

%\author{ Belle Collaboration}
\ \\
% to make it single spaced
%\tighten

\begin{abstract}
\noindent
%{\bf Abstract:\ }
We 
report the results of a search
for a charmonium-like state produced in the process
$\gamma \gamma \to \omega J/\psi$ in the 3.9--4.2~GeV/$c^2$ mass region.
We observe a significant enhancement, which is well-described by 
a resonant shape with mass  
$M = (3915 \pm 3 \pm 2)~{\rm MeV}/c^2$ and total width
$\Gamma = (17 \pm 10 \pm 3)~{\rm MeV}$. 
This enhancement
may be related to one or more 
of the three charmonium-like states so far reported in the 
3.90--3.95~GeV/$c^2$ mass region.
\end{abstract}

%{\renewcommand{\thefootnote}{\fnsymbol{footnote}}

\maketitle
\tighten

\setcounter{footnote}{0}
\setcounter{figure}{0}

\normalsize

%%\section{Introduction}
 Many new charmonium-like states have been discovered
by the B-factory experiments, typically via a prominent
hadronic decay to a known charmonium state, 
such as $J/\psi$, $\psi(2S)$ or $\chi_{c1}$.
Some have attracted particular  interest because of 
their net electric charge~\cite{zplus}.
Three of the new (neutral) states 
were discovered by Belle in the 3.90--3.95~GeV/$c^2$ mass region. 
The $X(3940)$ was 
found in the  $e^+ e^- \to J/\psi X$
double charmonium production process,
with a prominent decay to the $D\bar{D}^*$
final state~\cite{bellex}. 
The $Y(3940)$ was
observed in the $B$ decay process
$B^- \to Y(3940) K^-$ with  $Y(3940) \to \omega J/\psi$~\cite{belley,babary},
and is a candidate for an exotic state, such as a hybrid meson ($c\bar{c}g$),
or a $D^*\bar{D}^*$ bound state~\cite{branz}. 
The $Z(3930)$ was 
found in the $\gamma \gamma \to D\bar{D}$ process~\cite{bellez},
and is usually identified with the
$\chi_{c2}(2P)$.  
These three states appear in different
production and decay processes, and are usually
considered to be distinct particles, however there is no
decisive evidence for this. 
The interpretation of
these states has been
discussed by many authors: see, {\it e.g.}, Ref.~\cite{godfrey}.

 It is important to search for a signature of the
$Y(3940)$ or any other resonant state contributing to
two-photon production of $\omega J/\psi$. 
This final state is the lightest combination of two vector mesons 
with definite $C$-even and $I=0$ quantum numbers that can be 
produced  in two-photon processes via a hidden-charm state. 
In this paper we present   
measurements of the
$\gamma \gamma \to \omega J/\psi$
process in the 3.9--4.2~GeV/$c^2$ mass region,
in which we observe a resonant enhancement. 
The signal is
from the two-photon process
$e^+e^-\rightarrow e^+e^- \omega J/\psi$
in the ``zero-tag'' mode,  where
neither the final-state electron nor positron 
recoiling from photon emission are detected.
%The $\omega J/\psi$ system which we exclusively detect 
%has very small transverse momentum with respect to the beam axis
%in this case.

%%\section{Detector and Data}
We use experimental data recorded with
the Belle detector~\cite{belle} at the KEKB $e^+e^-$ asymmetric-energy
(3.5 on 8 GeV) collider~\cite{kekb}, corresponding to
an integrated luminosity of 694~fb$^{-1}$.  
The data are accumulated mainly on the
$\Upsilon(4S)$ resonance $(\sqrt{s} = 10.58~{\rm GeV})$
and 60~MeV below it. A small fraction of 
data from different beam energies near 10.36~GeV (the 
$\Upsilon(3S)$ mass) and  10.87~GeV (the $\Upsilon(5S)$ mass)
is also included in the sample.
%The effect from the difference in the two-photon luminosity function at these
%different energy points is less than 0.5\% for the 
%whole sample, for the two-photon center-of-mass energy ($W$) 
%region below 4.3~GeV.

A comprehensive description of the Belle detector is
given elsewhere~\cite{belle}.
Charged tracks are reconstructed in a central
drift chamber (CDC) located in a uniform 1.5~T solenoidal magnetic field.
The $z$ axis of the detector and the solenoid is along the positron beam,
with the positrons moving in the $-z$ direction.
Track trajectory coordinates near the
collision point are measured by a
silicon vertex detector (SVD).  Photon detection and
energy measurements are provided by a CsI(Tl) electromagnetic
calorimeter (ECL). A combination of 
silica-aerogel Cherenkov counters (ACC), 
a time-of-flight counter (TOF) system consisting
of a barrel of 128 plastic scintillation counters,
and specific ionization ($dE/dx$) measurements in the CDC
provides $K/\pi$ separation for charged tracks over a wide
momentum range. The magnet return iron is instrumented to 
form a $K_L$ detection and muon 
identification (KLM) system that 
detects muon tracks.

Signal candidates are triggered by a variety
of track triggers that require two or more CDC
tracks associated with TOF hits, ECL clusters, a total energy 
deposit in the ECL above a threshold (0.5~GeV), or a muon track 
in the KLM detector. In addition, events with a total ECL energy
above 1.1~GeV are triggered by a separate logic.  
Because of the presence of a lepton pair in the final state
of the signal processes, a combination of the above triggers
provides a high overall
trigger efficiency, ($98\pm 2$)\%.
%%, with a complementary contribution from each component.

%%\section{Event Selection}

 We select signal event candidates by reconstructing all 
the final state particles
from $\omega \to \pi^+\pi^-\pi^0$ and 
$J/\psi \to l^+l^-$ ($l = e \ {\rm or}\  \mu)$.
Twelve selection criteria are imposed: 
(1) there are just 4 charged tracks
with transverse momentum $p_t > 0.1$~GeV/$c$ originating in
the beam collision region; (2) the net charge of the tracks is zero; 
(3) none of the tracks is identified as a kaon (we
require a likelihood ratio ${\cal L}(K)/({\cal L}(K)+{\cal L}(\pi)) < 0.8$,
which is satisfied by 99.6\% of pions but only 5\% of
kaons, for momenta below 0.8~GeV/$c$);
(4) there is a net-charge-zero combination of two tracks
whose invariant mass is in  the $J/\psi$ mass region,
$|M(2~{\rm tracks}) - M_{J/\psi}|<0.2$~GeV/$c^2$,
where $M_{J/\psi}=3.0969$~GeV/$c^2$ and we assume the pion mass for 
each of the two tracks;
(5) there is one or more neutral pion candidate formed by a 
mass-constrained fit of two photons, with $\chi^2 < 4$; 
(6) the number of $\pi^0$'s with $p_t > 0.1$~GeV/$c$ must not exceed 
one in an event. If there is a $\pi^0$ satisfying the $p_t$ condition, 
it is accepted as the only  $\pi^0$ candidate in that event. 
If no $\pi^0$ satisfies
the $p_t$ condition, we retain all the $\pi^0$ candidates 
(below $p_t<0.1$~GeV/$c$) at this stage;
(7) events in the kinematic region of
initial-state-radiation processes,  $e^+e^- \to \gamma X$,
where the photon is emitted very close to the direction of the
incident $e^-$ beam, are eliminated. 
%(Such events, in the case where the
%photon is emitted very close to the direction of the
%incident $e^-$ beam and escapes detection, pass the
%selection up to this point.) 
We reject events 
satisfying the following condition on 
the  $z$-component of the laboratory momentum 
for the final-state system $X$,         
$P_z < (M_5^2-49.0~{\rm GeV}^2/c^4)/14.0~{\rm GeV}/c^3+ 0.6~{\rm GeV}/c$,
where $P_z$ and $M_5$ are the  $z$-component of the momentum and  
the invariant mass of the system constructed from the four 
tracks and a neutral pion candidate, respectively.

For further analysis we examine only
events with $W<4.3$~GeV, 
where $W$ is defined by $W = M_5 - M(l^+l^-)
+ M_{J/\psi}$, using a refined two-lepton invariant mass ($M(l^+l^-)$)
based on the lepton flavor identified by the following criteria: 
(8) if either of the tracks is identified as an electron,
based on the ECL energy deposit, the tracks are
identified as $e^+e^-$. Otherwise, if either track is identified 
as a muon based on KLM information, the tracks are identified as
$\mu^+\mu^-$. An event that fails both tests is rejected.
If one or more photons with energy between 20 and 200 MeV
are found within $3^\circ$ of either the $e^+$ or $e^-$ track, 
the energy of the most energetic photon near the track is 
added to the track momentum. 

Following this correction,
(9) we refine the $J/\psi$ selection with a more stringent requirement for
the lepton-pair invariant mass, 3.07~GeV/$c^2 < M(l^+l^-) < 3.12$~GeV/$c^2$;
(10) we suppress $\psi(2S) \pi^0$ events 
%which decay to the same final state as the signal process, 
with the mass difference requirement, 
$|M(l^+l^-\pi^+\pi^-) - M(l^+l^-) - 0.589~{\rm GeV}/c^2| > 0.01$~GeV/$c^2$; 
(11) to select $\omega$ candidate, a condition on the $\pi^+\pi^-\pi^0$ 
invariant mass, 0.753~GeV/$c^2 < M(3\pi) < 0.813$~GeV/$c^2$,
is imposed. 
If there are
multiple $\omega$ candidates due to
multiple $\pi^0$'s in an event, we choose the one with 
the smallest $\chi^2$ in the $\pi^0$ mass constrained fit,
in order to avoid multiple entries in the 
final $\omega J/\psi$ mass spectrum.
Finally, (12) we require transverse momentum balance 
for the 5-body system, $|\sum \mbox{\boldmath$p$}_t^*| < 0.1$~GeV/$c$, 
where  $\mbox{\boldmath$p$}_t^*$ is the momentum of a particle 
in the $e^+e^-$ c.m. frame, in the plane perpendicular to the 
beam direction~\cite{foot1}.  
%A small fraction of the data  (87~fb$^{-1}$), from early in the 
%experimental run, has been recovered from the event
%selection for a previous study. The following requirements, unrelated
%to the present analysis, were applied: 
%(A) there is no photon with energy exceeding 0.4~GeV in the event; 
%(B) the transverse momentum balance composed by the four charged 
%tracks is less than 0.2~GeV/$c$. These conditions reduce the selection
%efficiency for the signal process to 50--60\% of that in the main
%sample. 

Figures 1(a) and 1(b) show the distributions of $M(l^+l^-)$ just after 
requirement (8) and the mass difference $M(l^+l^-\pi^+\pi^-)- M(l^+l^-)$
just after requirement (9), respectively. 
%%We find a clear $J/\psi$ 
%%signal, some of which is due to $\psi(2S) \to \pi^+ \pi^- J/\psi$ 
%%events (see Fig.~1(b));
The $\psi(2S)$ contribution is effectively removed by criterion (10) above.

\begin{figure}
 \centering
   {\epsfig{file=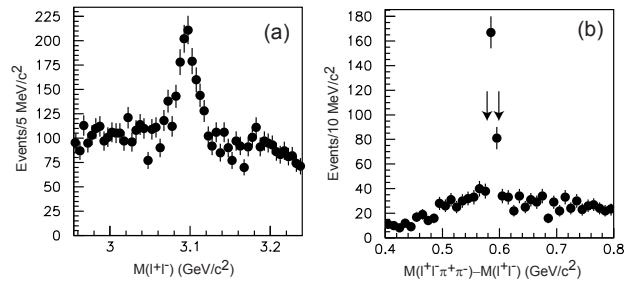,width=82mm}}
 \caption{(a) The $M(l^+l^-)$ distribution just after the dilepton
selection.
(b) The $M(l^+l^-\pi^+\pi^-)-M(l^+l^-)$ distribution after the
tight $J/\psi$ selection. Events between
the arrows are rejected as consistent with $\psi(2S)$
production.}

\label{fig:csrat}
\end{figure}

%%\section{Signal candidates and Background studies}
The main background process is multi-pion production from 
two-photon processes. However, after all selection
requirements are applied, 
non-$\omega J/\psi$ backgrounds are
rather small, as shown in the scatter plot in Fig.~2(a)
for the samples where all the selections except those for $M(l^+l^-)$
and $M(3\pi)$ are applied.
Figures~2(b) and 2(c) show the distributions of $M(l^+l^-)$ and $M(3\pi)$,
respectively, in the selection bands for the opposite-side 
particle; for clarity, we exclude 
events somewhat below the $\omega J/\psi$ 
threshold, $W \leq 3.85$~GeV. 
%Entries due to
%multiple $\omega$ candidates in a given event are 
%kept.

In Figs.~2(b) and 2(c),
the experimental $M(l^+l^-)$ and $M(3\pi)$ distributions 
%for the signal candidates of $J/\psi$ and $\omega$, respectively, 
are compared with those from the signal Monte Carlo (MC) events,
%In the figures, the signal MC events 
which are generated  
assuming spin-parity ($J^P$) and mass ($W$) of the $\omega J/\psi$ 
system to be $0^+$ and 3.93~GeV/$c^2$, respectively. 
Details of the signal MC generation are given below.
We confirm that the experimental mass distributions are 
consistent with those of signal MC events.

We find that there are two events in the signal region
with multiple $\omega$ candidates, out of 73 events in total; 
we choose only one combination in each event, according to 
criterion (11). The fraction is consistent with 
the 1--2\% 
multiple candidate rate
expected from the signal MC sample. 

% taking into account
%a small $W$ dependence in the $M(l^+l^-)$ resolution.
%By inspecting the distribution of the mass difference
%$M(l^+l^-\pi^+\pi^-)-M(l^+l^-)$ for the final candidates, 
%we confirm that the background from $\psi(2S)$ is 
%negligibly small.

We show the $W$ distribution  for the final 
$\gamma \gamma \to \omega J/\psi$ candidate events in Fig.~3.
There is a prominent resonance-like peak around 3.92~GeV.
It is far above the non-$\omega J/\psi$ background contribution,
which is 
estimated from the events in the $\omega$ and $J/\psi$ mass sidebands
(shown as shaded histograms for comparison); we define eight sideband regions 
in the plane of Fig.~2(a) with the same dimensions as the signal 
region, {\it i.e.}, each region centered at 3.035~GeV, 3.095~GeV 
or 3.155~GeV with a width of 0.05~GeV in the $M(l^+l^-)$ direction 
and centered at 0.693~GeV, 0.783~GeV and 0.873~GeV with a width of 
0.06~GeV in the $M(3\pi)$ direction, and average
the distribution over the eight regions. 
We modify the $W$ value of each sideband event plotted in Fig.~3,  
shifting it by the
difference between the sum of mass coordinates
of the central point of the signal region 
(3.878~GeV) from that of 
the sideband region where the event is found, 
for comparison to the signal-event distribution.
%Based on this sample, candidate events with
%$W>4.1$~GeV are dominated by 
%non-$\omega J/\psi$ backgrounds. 

Figure~4(a) shows a scatter plot of the transverse momentum balance 
{\it vs.} $W$ after requirement (11). A prominent concentration of events near 
$W=3.89-3.95$~GeV 
and $|\sum \mbox{\boldmath$p$}_t^*| < 0.05$~GeV/$c$ is visible; 
a comparison of the  $|\sum \mbox{\boldmath$p$}_t^*|$ 
projection with signal MC is shown in Fig.~4(b).
Based on these results, and the shape in $W$ (Fig.~3), we 
conclude that
the concentration of events is due to a resonance formed in
two-photon collisions.

\begin{figure}
 \centering
   {\epsfig{file=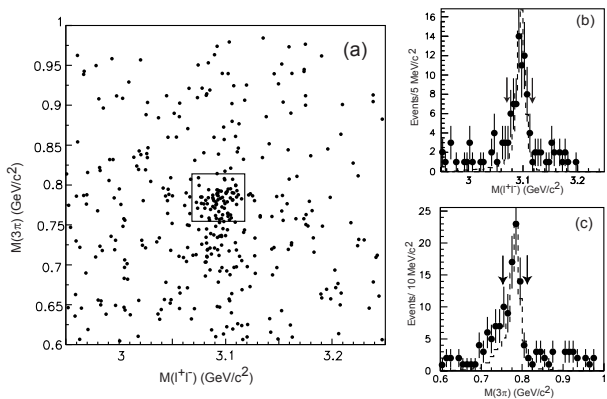,width=80mm}}
 \caption{(a) Scatter plot of $M(3\pi)$ {\it vs.} $M(l^+l^-)$ for the experimental
data, with all the other cuts applied. 
The rectangle shows the signal region.
%(b) The $M(l^+l^-)$ distribution for the experimental
%event samples, with all the other cuts applied. 
%(c) The $M(3\pi)$ distribution 
%for the experimental
%event samples all the other cuts applied. 
(b) The $M(l^+l^-)$ and (c) $M(3\pi)$ distributions
%for the experimental event samples, 
with all the other cuts applied, 
and requiring $W > 3.85$~GeV. Entries 
due to multiple $\omega$ candidates in a given event are 
included.
%(c) distribution 
%for the experimental
%event samples all the other cuts applied. 
%An additional cut,
%$W > 3.85$~GeV, is applied for (b) and (c).
%In (b) and (c), 
Points with error bars show the data;
the dashed histograms show signal MC events, 
normalized to the data yield in the selection area, 
with the scaled 
sideband yield
subtracted. The arrows show the selection regions.}
\label{fig:csrat}
\end{figure}

%%\section{Derivation of the resonance parameters}
The $W$ distribution for the final candidate events is 
fitted by an incoherent sum of resonant 
and background components. We adopt an S-wave Breit-Wigner function 
with a variable width for the resonant component, 
$(2N_R/\pi)M^2\Gamma'/\{(W^2 - M^2)^2 + M^2\Gamma'^2)\}$ and 
$\Gamma' = \Gamma (p^*/p_0^*)$, 
where $p^*$ is the momentum of the
two-body decay to $\omega J/\psi$, in the rest frame of a parent
particle of mass $W$;
$p^*_0$ is the value for $W=M$~\cite{belley}. 
The nominal mass ($M$),
width ($\Gamma$) and yield parameter ($N_R$) are treated
as fit parameters.

We represent the background
component by a quadratic function of $p^*$ that  
vanishes at the nominal $\omega J/\psi$  threshold, 
$M_{\rm th}=3.8796$~GeV/$c^2$.
We also add a constant term, to represent the 
high $W$ tail, which, as the sideband study suggests, is
dominated by non-$\omega J/\psi$ events.
The sum of the two components has a functional form,
$\{(a p^* + b p^{*2}) + c\}\theta(W-M_{\rm th})$, where $\theta(x)$ is 
a unit step function 
that is non-zero
only for $x>0$.
The parameters $a$, $b$ and $c$ are floated within
the constraint that each of the two background components
must be non-negative throughout the fitting region. 
%The sideband distributions are not included in the fit.

 The fit takes into account the $W$ resolution in the 
measurement, which is approximated by a double-Gaussian
function from the signal MC events (59\% of the signal 
has a resolution $\sigma$ of 4.5~MeV, while the remainder
has $\sigma=16$~MeV with the peak position displaced by 
$-4$~MeV).  We perform an unbinned maximum likelihood fit 
in the region 3.875~GeV$<W<4.2~$GeV. 
The signal candidates with the smallest $W$ are the two
events with $W$ between 3.879 and 3.880~GeV.

 The $W$ dependences of the efficiency and luminosity 
function are taken into account in the fitting function. 
The efficiency is determined using signal MC events 
as described in detail later.
We use the $W$ dependence of the efficiency for $J^P=0^+$ for the 
nominal fit. Between the threshold and 3.96 GeV, the 
$W$-dependence is weak: the efficiency varies by 10\% only, 
and has a minimum near $W=3.92$~GeV.

The obtained resonance parameters for the mass and the 
width are as follows:
\begin{eqnarray}
M &=& (3915 \pm 3 \pm 2)~{\rm MeV}/c^2, \nonumber \\
\Gamma &=& (17 \pm 10 \pm 3)~{\rm MeV}, \nonumber
\end{eqnarray}

\noindent
where the first and second errors are statistical and systematic,
respectively.
The estimated yield from the
resonant component in the fit is $49 \pm 14 \pm 4$ events 
in the region below 4.2~GeV.
The statistical significance of the resonant peak is  $7.7 \sigma$,
which is determined
from the difference of the logarithmic likelihoods, 
$-2 {\rm ln}(L/L_0)$,
taking the difference of the number of degrees of freedom 
in the fits into account,
where $L_0$ and $L$ are the likelihoods of the fits
with and without a resonant component, respectively.
The relevant fit curves are shown in Fig.~3.
The $\chi^2$ of the nominal fit, determined using
10-MeV-width binning in the range 3.85--4.2~GeV, is
27.3, for 29 degrees of freedom.

The systematic errors quoted above are 
determined from a study of 
alternate fits: we use
a Breit-Wigner function with a constant width; 
we enlarge the invariant-mass resolution by 20\% 
(an over-estimate of the data-Monte Carlo
difference allowed by the fit); 
we change the upper limit of the fit region in $W$ 
to 4.1~GeV and 4.3~GeV, respectively.
%; we take the energy 
%dependence of efficiency for the $J^P=2^+$ assumption. 
The 
changes in the
central values of the 
corresponding 
resonance parameter are 
combined in quadrature.
We also take into account the uncertainty of the mass scale,
estimated to be 1~MeV/$c^2$, in the measurement of $M$.
There is no significant change in the parameters if 
$J^P=2^+$
is assumed; the changes 
of mass and width are less than 0.1~MeV/$c^2$ and 0.3~MeV,
respectively. The resonant contribution for the 
$J^P=2^+$ assumption is 1.0~event smaller than  
that for 
$J^P=0^+$.

\begin{figure}
 \centering
   {\epsfig{file=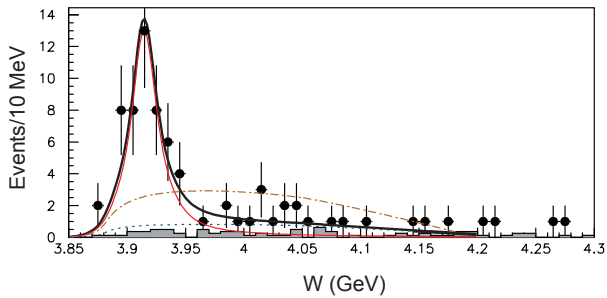,width=80mm}}
 \caption{The $W$ distribution of the final candidate events 
(dots with error bars). 
The shaded histogram is the distribution of non-$\omega J/\psi$
backgrounds estimated from the sideband distributions. The bold solid,
thinner solid and dashed curves 
are the total, resonance and background
contributions, respectively, from the standard fit (see the text). 
The dot-dashed curve is the fit without a resonance.}
%(black), thinner solid (dark red) and dashed (blue) curves 
%are the total, resonance and background
%contributions, respectively, from the standard fit 
%(see the text). The dot-dashed (orange-color)
%curve is the fit without assuming a resonance.}
\label{fig:csrat}
\end{figure}

\begin{figure}
 \centering
   {\epsfig{file=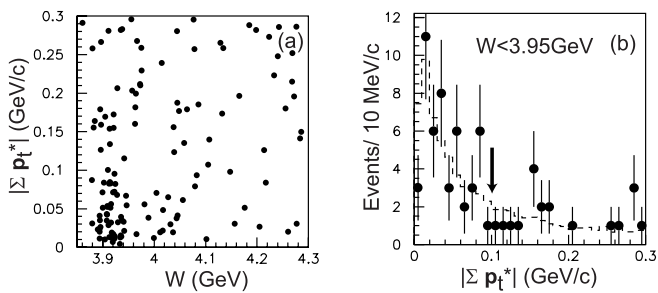,width=87mm}}
 \caption{(a) Scatter plot of $p_t$ balance {\it vs.} $W$ for the final 
candidate events in which only requirement (12) is omitted. 
(b) The projection onto the $p_t$ balance axis for 
events with $W<3.95$~GeV. The dashed histogram is the expectation
from signal MC events, normalized to the number 
of signal candidates in the selected region. 
The $p_t$ balance requirement is indicated by the arrow.
}
\label{fig:csrat}
\end{figure}
%%\begin{center}
\begin{table}[t]
\caption{Sources and sizes of systematic error in the 
efficiency determination}
\label{tab:systerr}
\begin{tabular}{c|c} \hline \hline
Source  & Syst.~ error (\%) \\
\hline
Trigger efficiency & 2 \\
Track reconstruction & 4 \\
$\pi^0$ reconstruction & 3 \\
Particle identification cuts & 2 \\
Effect of background hits & 3 \\
$J/\psi$ selection & 3 \\
$\omega$ selection & 6 \\
$W$-dependence, effect of background, etc. & 3 \\
Luminosity function, integrated luminosity & 5 \\
\hline
Total & 11\% \\
\hline \hline
\end{tabular}
\end{table}
%%\end{center}
%%\section{Efficiency and its systematic error}
The efficiency for selecting $\gamma \gamma \to \omega J/\psi$
events is determined using signal MC events generated by TREPS
code~\cite{treps}. We generate $10^5$ MC events for both 
$e^+e^-$ and $\mu^+\mu^-$ decays of $J/\psi$, at  
nine different $W$ points between 3.89 and 4.15~GeV.
The efficiency for the signal process at $W=3.92$~GeV 
is determined to be $(1.85 \pm 0.20)\%$ ($(1.26 \pm 0.14)\%$)
for the $J^P=0^+$ ($2^+$) assumption; 
the efficiency is 
defined for the range 
%of the virtuality of each incident photon 
$Q^2 < 1.0$~GeV$^2$, for each incident photon. 
We assume
production in helicity-2 for $J^P=2^+$~\cite{bellez} and
decay to $\omega J/\psi$ in an S-wave for both $0^+$ and $2^+$.
The other 
two possible $J^P$ assumptions, $0^-$ and $2^-$, 
are similar and 
give an efficiency close to that for $J^P=0^+$.

Based on the efficiencies calculated for the two $J/\psi$ decay
modes, 
the fraction of signal in the $e^+e^-$ mode is expected
to be 36\%. This is consistent with the fraction in the data:
27 $J/\psi \to e^+e^-$ events among 
the 73 signal candidates.

Sources of systematic errors in the efficiency determination
and their contributions are listed in Table I.
We confirm that the inefficiency due to each of the particle identification
cuts, (3) and (8), is very small, less than 1\%, for signal events.
The uncertainties in the efficiencies of the invariant mass cuts
are estimated by  varying the selection regions near
$M(l^+l^-) = M_{J/\psi}$ and $M(3\pi) = M_{\omega}$
by $\pm 20\%$ in the MC.  
We sum the uncertainties in quadrature, and find 11\% in total.
%The total systematic error on the detection efficiency is 11\%
%from their quadratic sum.

%%\section{Discussion and Conclusion}
Treating the observed structure as 
a resonance 
denoted by
$X(3915)$, we derive the product
of the two-photon decay width and the branching fraction 
to $\omega J/\psi$, using the yield parameter $N_R$ from
the fit and the selection efficiency. We obtain
\begin{eqnarray}
\Gamma_{\gamma \gamma}(X(3915)) {\cal B}(X(3915) \to \omega J/\psi)\ \ \ \ \ \  \ \ \ \ \ \ \ \ \ \ \ \ \ \ \ \ \nonumber \\  
\ \ \ \ \ \ \ \ \ \ \ = \left\{ \begin{array}{ll}
(61 \pm 17 \pm 8)~ {\rm eV} & \mbox{for $J^P=0^+$} \\
(18 \pm 5 \pm 2)~ {\rm eV} & \mbox{for $J^P=2^+$, helicity-2.} 
\end{array} \right. \nonumber
\end{eqnarray}
Based on this result, and the measured width $\Gamma$,
the product of the two partial widths of the $X(3915)$,
$\Gamma_{\gamma\gamma}(X) \Gamma_{\omega J/\psi}(X)$ is of
order $10^3$~keV$^2$. If we assume $\Gamma_{\gamma\gamma}
\sim {\cal O}(1$~keV), typical for an excited charmonium state,
this implies $\Gamma_{\omega J/\psi} \sim {\cal O}(1$~MeV): a
rather large value for a charmonium-transition
%even for a charmonium-inclusive
partial width of such a state. 
%Prediction of the partial decay 
%widths of $Y(3940)$ based on a $D^*\bar{D}^*$ bound-state 
%model~\cite{branz} derives their product roughly compatible 
%to the present result.
This value of the product of the partial decay widths is 
roughly compatible with the prediction assuming the 
$D^*\bar{D}^*$ bound-state model~\cite{branz}. 

To conclude, we have observed a resonance-like 
enhancement in the $\gamma \gamma
\to \omega J/\psi$ process with a statistical significance of
$7.7\sigma$, which contains 
$49 \pm 14 \pm 4$
events in the peak component. 
The mass and width have been measured to be 
$M = (3915 \pm 3 \pm 2)~{\rm MeV}/c^2$ and
$\Gamma = (17 \pm 10 \pm 3)~{\rm MeV}$, respectively.
These values are consistent with
those of 
the $Y(3940)$, which is seen in the $\omega J/\psi$ final
state~\cite{belley, babary}, and close to those of the
$Z(3930)$, which is seen in $\gamma \gamma \to D\bar{D}$~\cite{bellez}.  

%{\color{red}
%which has been observed in a $B$-meson decay}
%process~\cite{belley, babary}, and those of the $Z(3930)$,  
%{\color{red} 
%which has been observed in the decay to $D\bar{D}$}
%final state~\cite{bellez}. 

%%{\bf Acknowledgements}
%-------- Short version, if necessary, for PRL -----------
We thank the KEKB group for excellent operation of the
accelerator, the KEK cryogenics group for efficient solenoid
operations, and the KEK computer group and
the NII for valuable computing and SINET3 network support.  
We acknowledge support from MEXT, JSPS and Nagoya's TLPRC (Japan);
ARC and DIISR (Australia); NSFC (China); 
DST (India); MEST, KOSEF, KRF (Korea); MNiSW (Poland); 
MES and RFAAE (Russia); ARRS (Slovenia); SNSF (Switzerland); 
NSC and MOE (Taiwan); and DOE (USA).


\begin{thebibliography}{99}
\bibitem{foot1}
 
A small fraction of the data  (87~fb$^{-1}$), from early in the 
experimental run, has been recovered from the event
selection for a previous study~\cite{bellez}.
The following requirements, unrelated
to the present analysis, were applied: 
(A) there is no photon with energy exceeding 0.4~GeV in the event; 
(B) the net transverse momentum of the four charged 
tracks is less than 0.2~GeV/$c$. These conditions reduce the selection
efficiency for the signal process to 50--60\% of that in the main
sample.
\end{thebibliography}
\end{document}